\documentclass[pra,12pt,eqsecnum,showpacs]{revtex4}
\usepackage{epsf}
\usepackage{graphicx}

\begin{document}

 \def\BE{\begin{equation}}
 \def\EE{\end{equation}}
 \def\BEA{\begin{eqnarray}}
 \def\EEA{\end{eqnarray}}

\title{Quantum limits of super-resolution in reconstruction of optical objects}
\author{Vladislav N.~Beskrovnyy}%
\author{Mikhail I.~Kolobov}%
\affiliation{ Laboratoire PhLAM, Universit\'e de Lille 1, F-59655,%
Villeneuve d'Ascq Cedex, France}%

\pacs{42.50.Dv, 42.30.Wb, 42.50.Lc}

\date{\today}

\begin{abstract}
We investigate analytically and numerically the role of quantum
fluctuations in reconstruction of optical objects from diffraction-limited
images. Taking as example of an input object two closely spaced Gaussian
peaks we demonstrate that one can improve the resolution in the
reconstructed object over the classical Rayleigh limit. We show that the
ultimate quantum limit of resolution in such reconstruction procedure is
determined not by diffraction but by the signal-to-noise ratio in the
input object. We formulate a quantitative measure of super-resolution in
terms of the optical point-spread function of the system.
\end{abstract}

\maketitle

\section{Introduction}

Quantum imaging is a new branch of quantum optics that investigates the
ultimate performances of optical imaging allowed by the laws of quantum
mechanics~\cite{Kolobov99,Lugiato02}. One of the pertinent questions of
quantum imaging is about the quantum limits of the optical resolution. The
classical resolution limit was established by Abbe and Rayleigh at the end
of the nineteen century and is very well known nowadays. The Rayleigh
resolution criterion states that the resolution in an optical system is
limited by diffraction on its optical elements due to the wave nature of
the light. According to this criterion, two closely spaced points at the
input of the optical system cannot be resolved if the distance between
them is smaller than $\lambda/{\rm NA}$ where $\lambda$ is the wavelength
of the light and ${\rm NA}$ - the numerical aperture of the optical
system.

This classical resolution limit  is based on the presumed resolving
capabilities of the human eye and is not a fundamental limit like, for
example, the Heisenberg uncertainty relation. Nowadays using modern CCD
cameras for detection of optical images with subsequent electronic
processing one can often improve the resolution beyond the classical
diffraction limit. For example, while the Rayleigh criterion puts the
resolution limit of about 0.2 $\mu$m for an optical microscope, by
processing microscopic images one can achieve a precision of about
1nm~\cite{Kamimura87}. Even more spectacular results has been obtained
recently for detection of a small displacement of a laser
beam~\cite{Treps03}. Using "spatially squeezed" laser beam the authors
of~\cite{Treps03} have succeeded to measure a transverse displacement of
1.6 Angstrom of a laser beam with wavelength $\lambda=1064$ nm.

These two experiments are particular examples of the so-called
super-resolution techniques that aim to improve the optical resolution
beyond the classical diffraction limit. Super-resolution is possible when
one has some {\it a priori} information about the object. For example, in
practice one usually deals with objects of finite size. In this case one
can obtain super-resolution because the spatial Fourier spectrum of such
objects, produced in the focal plane of a lens where the system pupil is
located, is an analytical function. Thus, measuring a part of the Fourier
spectrum, transmitted by the system pupil, one can in principle
reconstruct the rest of the spectrum via the analytical continuation and,
therefore, obtain an infinite resolution. However, the procedure of such
an analytical continuation is extremely sensitive to different kind of
noises present in the optical system. Recently it was
shown~\cite{Kolobov00} that the ultimate limit of resolution in
diffraction-limited imaging is determined not by diffraction but by the
quantum fluctuations of light within the object area and the vacuum
fluctuations outside it. These quantum fluctuations set up the {\it
standard quantum limit of resolution} which can be by the orders of
magnitude smaller than the diffraction limit. Moreover, one can further
improve the resolution beyond the standard quantum limit using multimode
squeezed light. An optical scheme of such a super-resolution microscopy
with squeezed light was proposed in Ref.~\cite{Sokolov04}.

In this paper we numerically simulate the role of quantum fluctuations in
reconstruction of optical objects from the diffraction-limited images. We
take as example of an input object two Gaussian peaks placed so close to
each other that in the output image they cannot be resolved according to
the Rayleigh criterion. We assume that one can measure the spatial Fourier
components of the input object in the focal plane of the imaging lens
within the pupil area. This kind of measurement corresponds to the
super-resolving Fourier-microscopy~\cite{Scotto03} since instead of
observing the image one observes its spatial Fourier spectrum. We
formulate the quantum theory of this microscopy in terms of the prolate
spheroidal functions very similar to the theory developed in
Ref.~\cite{Kolobov00}. Using this quantum theory, we simulate numerically
the reconstructed objects and investigate the role of quantum fluctuations
on their resolution.

Our numerical simulations allow us to confirm that when quantum
fluctuations are not taken into account, one can easily improve the
resolution in the reconstructed objects over the diffraction limit by one
order of magnitude. When the object is illuminated by a light wave in
coherent state, the quantum fluctuations of light inside the object and
vacuum fluctuations outside it set up the standard quantum limit of
resolution which depends on the signal-to-noise ratio in the input object.
Finally, we demonstrate that one can go beyond the standard quantum limit
of resolution using the multimode squeezed light for illumination of the
object.

The paper is organized as follows. In Section~\ref{QuantumTheory} we
formulate the quantum theory of the super-resolving Fourier-microscopy in
terms of prolate spheroidal wave functions. In
Section~\ref{QuantumFluctuations} we give numerical examples of
super-resolution of the input object consisting of two closely spaced
Gaussian peaks and illustrate qualitatively the role of quantum
fluctuations on the degree of super-resolution. In
Section~\ref{Point-spread-function} we provide quantitative measure of
super-resolution in terms of the optical point-spread function. In
particular we plot super-resolution factor as a function of the mean
photon number in the input object for the coherent light and multimode
squeezed light. We discuss our results and the future perspectives in
Section~\ref{Discussions}.

\section{Quantum theory of the super-resolving Fourier-microscopy}
\label{QuantumTheory}

In this section we shall present the original quantum theory of the
super-resolving Fourier-microscopy in terms of prolate spheroidal
functions. The optical scheme of diffraction-limited coherent optical
imaging is shown in Fig.~\ref{scheme}. For simplicity we consider
one-dimensional case. The object of finite size $X$ is placed in the
object plane. The first lens $L_1$ performs the spatial Fourier transform
of the object into the pupil plane with a pupil of finite size $d$.
Diffraction on this pupil is a physical origin of the finite resolution in
our scheme (we neglect diffraction on the imaging lenses). The second lens
$L_2$ performs the inverse Fourier transform and creates a
diffraction-limited image in the image plane.

As mentioned above, to achieve super-resolution one needs some {\it a
priori} information about the object. In our case we know {\it a priori}
that the object is confined within the area of size $X$ and is identically
zero outside. The spatial Fourier transform of such an object is an entire
analytical function. Therefore, knowing the part of the Fourier spectrum
within the area $d$ of the pupil allows for an analytical continuation of
the total spectrum and, therefore, for unlimited resolution. However, this
analytical continuation is extremely sensitive to the noise in the
diffracted image and, as shown in Ref.~\cite{Kolobov00}, is limited by the
quantum fluctuations of light.

Quantum theory of the diffraction-limited optical imaging in
Fig.~\ref{scheme} was developed in Ref.~\cite{Kolobov00} in terms of
prolate spheroidal wave functions. These are the eigenfunctions of the
imaging operator describing the transformation of the optical field from
the object plane into the image plane. One can decompose the input object
and the output image over these eigenfunctions and obtain the relation
between the decomposition coefficients. Then detecting the output image
with, for example, a sensitive CCD camera, one can evaluate the
decomposition coefficients of the image. Using the relation between the
decomposition coefficients of the image and the object one can reconstruct
the latter with resolution better than the classical diffraction limit.

Our numerical simulations in Ref.~\cite{Kolobov03} have shown, however,
that for evaluation of the decomposition coefficients one has to detect
the output image over unrealistically large area in the image plane due to
the oscillating behavior of the prolate functions. That is why in
Ref.~\cite{Scotto03} we have proposed a modified version of the scheme
where the CCD camera is placed in the pupil plane instead of the image
plane, and one detects the spatial Fourier spectrum. We have called this
modified scheme super-resolving Fourier-microscopy. The advantage of this
set-up is that now the spatial Fourier spectrum is measured over the
finite region within the pupil. To understand the role of the quantum
fluctuations on the resolution of the Fourier-microscopy we need to
formulate the quantum theory of this modified scheme.

Let us introduce the dimensionless spatial coordinates in the object plane
as $s = 2x/X$, and in the pupil plane as $\xi=2y/d$ (see Fig.~1). The
dimensionless photon annihilation operators in the object plane will be
denoted as ${\hat a}(s)$ and in the pupil plane as ${\hat f}(\xi)$. These
operators obey the standard commutation relations,
 \begin{equation}
      [{\hat a}(s),{\hat a}^\dag(s')] = \delta(s-s'), \quad
      [{\hat f}(\xi),{\hat f}^\dag(\xi')] = \delta(\xi-\xi'),
           \label{comm}
 \end{equation}
and are normalized so that $\langle {\hat a^{\dag}}(s){\hat a}(s)\rangle$
gives the mean photon number per unit dimensionless length in the object
plane and $\langle {\hat f^{\dag}}(\xi){\hat f}(\xi)\rangle$ - in the
pupil plane. The spatial Fourier transform $(T{\hat a})(\xi)$ performed by
the lens $L_1$, in terms of these dimensionless variables reads as
follows,
 \begin{equation}
    {\hat f}(\xi) = (T{\hat a})(\xi) =
    \sqrt{\frac{c}{2\pi}}\int_{-\infty}^\infty {\hat a}(s) e^{-i
    cs\xi}ds,
         \label{Fourier}
 \end{equation}
where $\displaystyle{c=\frac{\pi}{2}\frac{dX}{\lambda f}}$ is the
space-bandwidth product of the imaging system. It is important to note
that the limits of integration in this equation are over the whole object
plane since it concerns operators and not the classical $c$-numbers.

As in Ref.~\cite{Kolobov00} we shall formulate our quantum theory of
super-resolving Fourier-microscopy in terms of prolate spheroidal
functions $\psi_k(s)$\cite{Slepian61,Frieden71}. These are the
eigenfunction of the imaging operator of the scheme, orthonormal on the
interval $-\infty <s<\infty$. To obtain the canonical transformation of
the photon annihilation and creation operators from the object into the
pupil plane we shall split the coordinates $s$ and $\xi$ into two regions,
the "core", $|s|\leq 1$ and $|\xi|\leq 1$, corresponding to the area of
localization of the classical object and the transmission area of the
pupil, and the "wings", $|s|> 1$ and $|\xi|> 1$, outside these areas. The
orthonormal bases in these areas of the object plane are given by
 \BE
    \varphi_k(s) = \left\{\begin{array}{cl}
    {\displaystyle\frac{1}{\sqrt{\lambda_k}}\psi_k(s)} & \quad  |s|\leq 1,\\
    0 & \quad  |s|>1,\\
    \end{array} \right. \quad\quad\quad
    \chi_k(s) = \left\{\begin{array}{cl}
    0 & \quad |s|\leq 1,\\
    {\displaystyle\frac{1}{\sqrt{1-\lambda_k}}\psi_k(s)} & \quad  |s|>1,\\
    \end{array} \right.
           \label{varphi_chi}
 \EE
where $\lambda_k$ are the eigenvalues of the corresponding prolate
spheroidal functions $\psi_k(s)$, depending on the space-bandwidth product
$c$. We should note that the functions $\varphi_k(s)$ are complete in the
Hilbert space $L^2(-1,1)$. Similar relations take place in the pupil
plane.

In terms of two sets $\{\varphi_k(s)\}$ and $\{\chi_k(s)\}$ we can write
the annihilation operators in the object plane as
 \BE
     {\hat a}(s) = \sum_{k=0}^{\infty} {\hat a}_k \varphi_k(s) +
     \sum_{k=0}^{\infty} {\hat b}_k \chi_k(s),
              \label{object_plane}
 \EE
and in the pupil plane as
 \BE
     {\hat f}(\xi) = \sum_{k=0}^{\infty} {\hat f}_k \varphi_k(\xi) +
     \sum_{k=0}^{\infty} {\hat g}_k \chi_k(\xi).
              \label{pupil_plane}
 \EE
Here ${\hat a}_k$ and ${\hat f}_k$ are the annihilation operators of the
prolate modes $\varphi_k$ in the core region of the object and the pupil
planes, while ${\hat b}_k$ and ${\hat g}_k$ are the annihilation operators
of the prolate modes $\chi_k$ in the wings regions. The operators ${\hat
a}_k$ and ${\hat b}_k$ are expressed through the field operator ${\hat
a}(s)$ by
 \BE
  {\hat a}_k = \int_{-\infty}^\infty \,{\hat a}(s)\varphi_k(s)ds,
  \quad\quad
  {\hat b}_k = \int_{-\infty}^\infty \,{\hat a}(s)\chi_k(s)ds.
       \label{coefficients_core_wings}
 \EE
Similar relations hold for ${\hat f}_k$ and ${\hat g}_k$ and ${\hat
f}(\xi)$.

Four sets of operators ${\hat a}_k$, ${\hat b}_k$, ${\hat f}_k$, and
${\hat g}_k$ obey the standard commutation relations of the photon
annihilation and creation operators of the discrete modes. For example,
for the operators ${\hat a}_k$ in the core region of the object plane we
have,
 \BE
      [{\hat a}_k,{\hat a}_{k'}^\dag] = \delta_{kk'}, \quad
      [{\hat a}_k,{\hat a}_{k'}] = 0.
        \label{commutation}
 \EE
The same relations are fulfilled for three other groups of operators
${\hat b}_k$, ${\hat f}_k$, and ${\hat g}_k$. It is clear that creation
and annihilation operators of different groups are independent and commute
with each other.

In our analysis we shall use the following properties of prolate
spheroidal functions~\cite{Frieden71},
 \BE
 \int_{-1}^{1} \varphi_k(s) e^{-ics\xi}ds =
 (-i)^k\sqrt{\frac{2\pi}{c}} \psi_k(\xi),
       \label{prolate_property_1}
 \EE
 \BE
 \int_{-\infty}^{\infty} \psi_k(s) e^{-ics\xi}ds =
 (-i)^k\sqrt{\frac{2\pi}{c}} \varphi_k(\xi),
      \label{prolate_property_2}
 \EE
Using (\ref{varphi_chi}), the field transform (\ref{Fourier}) between the
object and the pupil plane and these properties, we find the following
propagation relations for the core and wings of the light wave,
 \BE
 \left(T\varphi_k\right)(\xi) = (-i)^k\left[\sqrt{\lambda_k}
 \varphi_k(s) + \sqrt{1-\lambda_k} \chi_k(s)\right],
       \label{propagation_1}
 \EE
 \BE
 \left(T\chi_k\right)(\xi) = (-i)^k\left[\sqrt{1-\lambda_k}
 \varphi_k(s) - \sqrt{\lambda_k} \chi_k(s)\right].
       \label{propagation_2}
 \EE
Substituting Eqs.~(\ref{object_plane}) and (\ref{pupil_plane}) into the
field transform (\ref{Fourier}) and using Eqs.~(\ref{propagation_1}) and
(\ref{propagation_2}), we arrive at the following relations between the
photon annihilation operators of the prolate modes in the object and the
pupil planes,
\begin{equation}
     \hat f_k =
     (-i)^k(\sqrt{\lambda_k} \hat a_k + \sqrt{1-\lambda_k} \hat b_k ),
        \label{quant_trans_core}
\end{equation}
\begin{equation}
     \hat g_k =
     (-i)^k(\sqrt{1-\lambda_k} \hat a_k - \sqrt{\lambda_k} \hat b_k ).
        \label{quant_trans_wings}
\end{equation}
This relations are similar to the transformation performed by a
beam-splitter with the amplitude transmission coefficients
$(-i)^k\sqrt{\lambda_k}$ and the reflection coefficients
$(-i)^k\sqrt{1-\lambda_k}$, and preserves the commutation relation of the
annihilation and creation operators in the pupil plane.

Let us assume that we can detect the spatial Fourier amplitudes ${\hat
f}(\xi)$ in the pupil plane within the transmission area of the pupil
using a sensitive CCD camera. This transmitted part of the spatial Fourier
spectrum is given by the first sum in Eq.~(\ref{pupil_plane}), the term
given by the second sum is absorbed by the opaque area of the pupil. It
should be emphasized that, since we need the complex field amplitudes and
not the intensities, one should use the homodyne detection scheme with a
local oscillator. Using Eqs.~(\ref{pupil_plane}),
(\ref{coefficients_core_wings}) and(\ref{quant_trans_core}) we can
calculate the operator-valued coefficients $\hat a_k ^{(r)}$ of the
reconstructed object as
\begin{equation}
     \hat a_k^{(r)} = \frac{\hat f_k}{(-i)^k\sqrt{\lambda_k}}=\hat a_k +
     \sqrt{\frac{1-\lambda_k}{\lambda_k}} \hat b_k,
          \label{quant-rest}
\end{equation}
where the superscript $(r)$ stands for "reconstructed". As follows from
Eq.~(\ref{quant-rest}), the reconstruction of the input object is not
exact because of the second term in Eq.~(\ref{quant-rest}). This term
contains the annihilation operators $\hat b_k$ responsible for the vacuum
fluctuations of the electromagnetic field in the area outside the object.
It is important to notice that these vacuum fluctuations prevent from
reconstruction of the higher and higher coefficients $\hat a_k$ in the
object because of the multiplicative factor
$\sqrt{(1-\lambda_k)/\lambda_k}$. Indeed, the eigenvalues $\lambda_k$
become rapidly very small after the index $k$ has attained some critical
value. This leads to rapid "amplification" of the vacuum fluctuations in
the reconstructed object that limits the number of the reconstructed
coefficients $\hat a_k$.

\section{Quantum fluctuations and reconstruction of the
spatial Fourier spectrum of the object}
\label{QuantumFluctuations}

\subsection{Reconstruction of classical noise-free objects}

In this section we shall illustrate numerically the role of quantum
fluctuations on the reconstruction of simple objects with super-resolution
beyond the classical diffraction limit. However, before taking into
account quantum fluctuations of light in the input object, we would like
to demonstrate the potential of the super-resolution technique with
prolate spheroidal functions for reconstruction of noise-free classical
objects, i.~e.~when the quantum fluctuations are neglected. This case
corresponds to the classical limit of the quantum theory developed in
previous section and can be simply obtained by taking mean values of the
operators.

In what follows we shall denote the classical complex amplitudes
corresponding to the quantum-mechanical operators by the same letters
without carets, for example $a(s)=\langle {\hat a}(s)\rangle$,
$a_k=\langle {\hat a}_k\rangle$, et cetera. Since the classical complex
amplitude of the object is zero outside the area $|s|\leq 1$, we have
$\langle {\hat b}_k\rangle=0$. Using Eq.~(\ref{object_plane}) we can write
this classical amplitude as
 \BE
     a(s) = \sum_{k=0}^{\infty} a_k \varphi_k(s).
              \label{object_plane_class}
 \EE
The classical complex amplitude $f(\xi)$ of the field in the pupil plane
is obtained from Eq.~(\ref{Fourier}),
\begin{equation}
    f(\xi) =
    \sqrt{\frac{c}{2\pi}}\int_{-1}^1 a(s) e^{-i
    cs\xi}ds,
         \label{Fourier_class}
 \end{equation}
with the integration limits over the object area, $|s|\leq 1$. Taking into
account the property of the prolate functions given by
Eq.~(\ref{prolate_property_1}), we can write the spatial Fourier spectrum
$f(\xi)$ as the following decomposition,
 \BE
     f(\xi) = \sum_{k=0}^{\infty}(-i)^k a_k \psi_k(s),
     \quad\quad -\infty<\xi<\infty.
              \label{pupil_plane_class}
 \EE
This spatial Fourier spectrum of the object spreads outside the
transmission area of the pupil $|\xi|\leq 1$. The spatial Fourier
components in the opaque area are absorbed and cannot be detected by the
CCD camera placed in the pupil plane. Super-resolution attempts to
reconstruct these absorbed Fourier components. From Eq.~(\ref{quant-rest})
we obtain the classical reconstructed coefficients,
 \begin{equation}
     a_k^{(r)} = a_k.
          \label{class-rest}
 \end{equation}
Because we have neglected the quantum fluctuations, the reconstructed
coefficients are identical to those of the input object. The classical
amplitude of the reconstructed object $a^{(r)}(s)$ can be written as the
following decomposition over the prolate functions,
 \BE
     a^{(r)}(s) = \sum_{k=0}^{L-1} a_k \varphi_k(s).
              \label{obj_reconst}
 \EE
Since in practice one can never have infinitely many coefficients $a_k$,
we have restricted the summation in this equation to $L$ first prolate
functions. When $L\rightarrow\infty$, the reconstructed object approaches
the exact one, $a^{(r)}(s)\rightarrow a(s)$. In practice the
super-resolution over the Rayleigh limit is determined by the number $L$
of terms used in the decomposition (\ref{obj_reconst}).

Alternatively to reconstruction of the object  itself one can try to
reconstruct its spatial Fourier spectrum as \BE
     f^{(r)}(\xi) = \sum_{k=0}^{L-1}(-i)^k a_k \psi_k(s),
     \quad\quad -\infty<\xi<\infty.
              \label{spectrum_reconst}
 \EE
Similarly to the reconstruction of the object, when $L\rightarrow\infty$,
the reconstructed spectrum approaches the exact one,
$f^{(r)}(\xi)\rightarrow f(\xi)$.

For numerical simulations we have taken a simple object of two narrow
Gaussian peaks,
\begin{equation}
     a(s)=  A\Bigl[\exp{\left(-\frac{(s-s_0)^2}{2\sigma^2}\right)} +
     \exp{\left(-\frac{(s+s_0)^2}{2\sigma^2}\right)}\Bigr],  \quad |s|\le 1,
     \label{eq:class-object}
\end{equation}
of width $\sigma$ separated by distance $2s_0$. We choose $2s_0=1$ and
$\sigma=0.1$, so that two peaks are well separated in the input object.
The normalization constant $A$ is chosen so that the integral of the
object intensity over the area of the object is equal to the total mean
number of photons $\langle {\hat N}\rangle$ in the object,
 \begin{equation}
    \int_{-1}^1 a^2(s) ds = \langle {\hat N}\rangle.
 \end{equation}
The Rayleigh resolution distance $R=\pi X/(2c)$ in dimensionless
coordinates $s$ is equal to $\pi/c$, where $c$ is the space-bandwidth
product. In our simulations we work with $c=1$. In this situation for
$2s_0<\pi$ we are beyond the Rayleigh limit.

In Fig.~2a we have plotted the normalized input object $a(s)/\sqrt{\langle
{\hat N}\rangle}$, in Fig.~2b - its spatial Fourier spectrum $f(\xi)$ in
the pupil plane, and in Fig.~2c - the output image $e(s)$ in the image
plane. Comparing the input object with its image one can clearly see that
it is impossible to resolve two Gaussian peaks in the image plane
according to the Rayleigh criterion. In Fig.~2b we have shown by the grey
color the opaque area of the pupil. The part of the spatial Fourier
spectrum of the object in this area is absorbed and therefore cannot be
detected by the CCD camera placed in the pupil plane. Below we shall
illustrate the reconstruction of these absorbed Fourier components by the
technique of the prolate functions.

For numerical simulations we had to evaluate two sets of prolate
spheroidal functions, $\varphi_k(s)$, defined on the interval $|s|\leq 1$,
and $\psi_k(s)$ defined for all $s$, $-\infty<s<\infty$. The first set is
necessary for decomposition of the input object $a(s)$, while the second
one is needed for reconstruction of the spatial Fourier spectrum
$f^{(r)}(\xi)$. For numerical calculations of $\varphi_k(s)$ we have used
the algorithm from Ref.~\cite{Xiao01}. In this algorithm the prolate
functions $\varphi_k(s)$ are evaluated as the series with the Legendre
polynomials $P_k(s)$,
 \begin{equation}
    \varphi_n(s)=\sum_{k=0}^{\infty} \gamma^{(n)}_k
    \sqrt{k+\frac{1}{2}}P_k(s), \quad\quad |s|\leq 1.
 \end{equation}
The coefficients $\gamma^{(n)}_k$ are found as the eigenvectors of the
symmetric matrix $A$ with the following nonzero elements,
 \begin{equation}
    A_{k,k} = k(k+1)+c^2\frac{2k(k+1)-1}{(2k+3)(2k-1)},
 \end{equation}
 \begin{equation}
    A_{k,k+2} = A_{k+2,k} = c^2\frac{(k+2)(k+1)}{(2k+3)\sqrt{(2k+1)(2k+5)}},
 \end{equation}
for all $k=0,1,2\dots$

We have written a numerical program in MATHEMATICA which implements this
algorithm. The advantage of this method is that it does not require the
direct solution of the eigenproblem for $\varphi_k(s)$ and $\lambda_k$
which is unstable due to the rapid decrease of the eigenvalues. This
algorithm allows us to calculate for $c=1$ at least $17$ first prolate
functions in spite of the fact that the eigenvalues of the higher order
functions become extremely small (for example, $\lambda_{17}=4.183\times
10^{-50}$). In Fig.~3 we show the first 17 prolate functions
$\varphi_k(s)$ evaluated by our numerical program.

For numerical calculation of the second set  of the prolate spheroidal
functions, $\psi_k(s)$, we have used the following property of the
Legendre polynomials ( see Eq.~(10.1.14) in Ref.~\cite{Abramowitz70}),
 \BE
 \int_{-1}^{1} P_n(s) e^{-ics\xi}ds =
 2i^n j_n(\xi),
 \EE
where $j_n(s)$ is the spherical Bessel function of the first
order~\cite{Abramowitz70}. Using this equation we can easily obtain the
following representation of $\psi_n(\xi)$,
 \begin{equation}
    \psi_n(\xi)=\sqrt{\frac{2c}{\pi}}i^n\sum_{k=0}^{\infty}(-i)^k \gamma^{(n)}_k
    \sqrt{k+\frac{1}{2}}j_k(c\xi), \quad\quad -\infty<\xi<\infty .
 \end{equation}
We illustrate the result of reconstruction of the spatial Fourier spectrum
of the input object in Fig.~4. In this figure we show the exact spatial
Fourier spectrum of the input object, drown by a solid line, as a function
of dimensionless coordinate $\xi$ in the pupil plane. Only part of this
spectrum within the transmission area of the pupil, $|\xi| \le 1$, is
transmitted to the image plane. The spatial Fourier harmonics in the
opaque area of the pupils, shown by grey color, are absorbed. This is a
reason of very large diffraction spread in the image plane shown in
Fig.~2c. The dashed lines in Fig.~4 correspond to the spatial Fourier
spectrum of the reconstructed object with $L=5, 7$ and 11 prolate
functions. One can see that the reconstructed spectrum approaches the
exact one for ever higher spatial frequencies $|\xi|$ as the number of
prolate functions increases.

With 7 prolate functions two spectra are very close to each other for
spatial frequencies $|\xi| \le 8$. This corresponds to a super-resolution
factor of 8 over the Rayleigh limit.

\subsection{Reconstruction of objects with quantum fluctuations}

For numerical simulations of quantum fluctuations we have chosen a
$c$-number representation of the quantum mechanical operators $\hat a_k$
and $\hat b_k$ in Eq.~(\ref{quant-rest}) corresponding to the antinormal
ordering of the creation and annihilation operators. In this
representation the operators $\hat a_k$ and $\hat b_k$ become the
$c$-number Gaussian stochastic variables $\alpha_k$ and $\beta_k$
respectively, which we shall write as
\begin{equation}
      \alpha_k=a_k+\delta \alpha_k, \quad\quad \beta_k=\delta \beta_k.
\end{equation}
Here $a_k=\langle \hat a_k\rangle$ is the mean value of the field
coefficients in the object area, and $\delta \alpha_k$ and $\delta
\beta_k$ are the stochastic Gaussian fluctuations. Note that the mean
values $\langle \hat b_k\rangle$ are zero because the classical field
component outside the object vanish. We have chosen the
antinormally-ordered representation because it remains valid even in the
case of the multimode squeezed state of light field at the input of the
scheme.

We introduce the quadrature components of the fluctuations $\delta
\alpha_k$ and $\delta \beta_k$ as follows,
\begin{equation}
      \delta \alpha_k=\delta X^{\alpha}_k+i\delta Y^{\alpha}_k, \quad\quad
      \delta \beta_k=\delta X^{\beta}_k+i\delta Y^{\beta}_k.
\end{equation}
When the input light is in the coherent state in the object area and in
the vacuum state outside, the correlation functions of the quadrature
fluctuations are equal to
\begin{equation}
      \langle\delta X^{\mu}_k \delta X^{\mu}_{k'}\rangle=
      \langle\delta Y^{\mu}_k \delta
      Y^{\mu}_{k'}\rangle=\frac{1}{4}\delta_{kk'},
\end{equation}
with $\mu=\alpha,\beta$.

If instead of coherent light we use the multimode squeezed light for
illumination of the object and multimode squeezed vacuum in the area
outside with subsequent homodyne detection at the pupil plane, these
correlation functions become
\begin{equation}
      \langle\delta X^{\mu}_k \delta X^{\mu}_{k'}\rangle=
      \frac{1}{4}e^{-2r}\delta_{kk'} \quad\quad
      \langle\delta Y^{\mu}_k \delta
      Y^{\mu}_{k'}\rangle=\frac{1}{4}e^{2r}\delta_{kk'},
\end{equation}
where $r$ is the squeezing parameter. In these formulas we have assumed
that the input light is amplitude-squeezed and for simplicity have chosen
the same squeezing parameter for all the essential modes that are used in
the decomposition of the reconstructed object.

The relative value of quantum fluctuations depends on the signal-to-noise
ratio in the input object which for the light in a coherent state is
determined by the total mean number of photons passed through the object
area during the observation time. For example, for a laser beam with
$\lambda=1064$ nm and optical power of 1 mW, and observation time of 1 ms
we obtain the mean photon number of $\langle {\hat N} \rangle=5.3\cdot
10^{12}$.

In Fig.~5a we have shown the results of reconstruction of the spatial
Fourier spectrum of the object from Fig.~\ref{fig:obj-img}a  when quantum
fluctuations of a coherent state are taken into account. The solid line
gives the exact spatial Fourier spectrum of the object. As in Fig.~4 the
grey area shows the absorbing part of the pupil. We use 7 prolate
functions and the mean photon number in the input object is taken $\langle
{\hat N} \rangle= 10^{12}$. The five thin lines correspond to the five
random Gaussian realizations of the quantum fluctuations in the coherent
state of $\hat a_k$ and the vacuum fluctuations of $\hat b_k$. The dashed
line corresponds to the reconstructed spectrum with 7 prolate functions
without noise. One can observe that the role of quantum fluctuations
becomes more and more important as one goes to the higher and higher
spatial frequencies where the random realizations of the Fourier spectra
deviate more and more from the mean value given by the dashed line.

In Fig.~5b we have increased the total mean value of photons to $\langle
{\hat N} \rangle= 10^{13}$. This corresponds to an increased
signal-to-noise ratio in the input object and should allow for better
super-resolution. This is illustrated in Fig.~5b by reduced deviation of
the random realizations from the mean value of the spectrum a compared to
Fig.~5a.

The same result can be achieved by using multimode squeezed light instead
of increasing the power of the source illuminating the object. This is
illustrated in Fig.~5c where we have used $\langle {\hat N} \rangle=
10^{12}$ as in the Fig.~5a, but have considered the light in a multimode
squeezed state with the squeezing parameter $e^r=10$ instead of the
coherent state. As the result the fluctuations in the higher spatial
frequencies are reduced that gives better super-resolution.

In next section we shall give a quantitative characteristic of
super-resolution as a function of the signal-to-noise ratio.

\section{Point-spread function for super-resolving reconstruction of objects}
\label{Point-spread-function}

In modern classical optics the resolution of an optical system is
characterized not by the two-point Rayleigh resolution criterion, but in
terms of its spatial transmission bandwidth. A typical optical system has
a finite band of spatial frequencies that are transmitted through the
system up to some cut-off frequency determined by the size of the system
pupil. The optical system is then said to be bandlimited or
diffraction-limited since diffraction effects on its pupil are responsible
for finite resolution.

A coherent diffraction-limited imaging system in classical optics can be
described by a linear equation relating the complex amplitude $a(s)$ of an
input object with the complex amplitude $e(s)$ of the
image~\cite{Bertero96},
 \begin{equation}
    e(s) = \int_{-\infty}^\infty h(s,s')a(s') ds',
         \label{imaging}
 \end{equation}
The impulse response function $h(s,s')$ that appears in this integral
equation represents the image at point $s$ in the image plane from a
point-source at point $s'$ in the object plane. For translationally
invariant or isoplanatic systems the impulse response depends only on the
difference $s-s'$ and the integral in (\ref{imaging}) becomes convolution,
 \begin{equation}
    e(s) = \int_{-\infty}^{\infty} h(s-s')a(s') ds'.
         \label{convolution}
 \end{equation}
In optics, the impulse response $h(s-s')$ is usually called {\it the
point-spread function} (PSF) of the system, and its Fourier transform {\it
the transfer function} (TF). For bandlimited optical systems the transfer
function is identically zero outside the transmission band of the system.
Super-resolution is defined as technique of restoring the spatial
frequencies of the object outside the transmission band~\cite{Bertero96}.
It is important to underline that in case when the object and the image
fields are related by the convolution ({\ref{convolution}),
super-resolution is impossible. To achieve super-resolution one needs some
{\it a priori} information about the input object. In our case the {\it a
priori} information is the assumption that the object has finite size.
With this assumption Eq.~(\ref{convolution}) in dimensionless coordinates
becomes~\cite{Kolobov00},
 \begin{equation}
    e(s) = \int_{-1}^1 h(s-s')a(s') ds',
         \label{finite_imaging}
 \end{equation}
with the imaging PSF $h(s-s')$ given by
 \begin{equation}
    h(s-s') = \frac {\sin[c(s-s')]}{\pi(s-s')}.
         \label{PSF}
 \end{equation}
For the reconstruction process we can write similar relation between the
reconstructed field operator ${\hat a}^{(r)}(s)$ and the object field
operator ${\hat a}(s)$. Using an operator-valued equivalent of
Eq.~(\ref{obj_reconst}) together with Eq.~(\ref{quant-rest}) we arrive at
the following result,
 \begin{equation}
     \hat a^{(r)}(s) = \int_{-1}^1 h^{(r)}(s,s')\hat a(s') ds'+
     \sum_{k=0}^{L-1}\sqrt{\frac{1-\lambda_k}{\lambda_k}} \hat
     b_k \varphi_k(s).
          \label{quant-rest-PSF}
 \end{equation}
Here the reconstruction point-spread function $h^{(r)}(s,s')$ is given by
 \begin{equation}
    h^{(r)}(s,s') = \sum_{k=0}^{L-1}\varphi_k(s)\varphi_k(s').
         \label{quant-PSF}
 \end{equation}
As seen from this equation, the form of the reconstruction PSF and, in
particularly, its width depends on the number of terms $L$ in the sum.
When this number grows infinitely, $L\to\infty$, the reconstruction PSF
tends to $\delta$-function,
 \begin{equation}
    \lim_{L\to\infty} h^{(r)}(s,s') = \sum_{k=0}^{\infty}\varphi_k(s)
    \varphi_k(s') = \delta(s-s'),
         \label{limit-quant-PSF}
 \end{equation}
and we have unlimited super-resolution. However, this ideal situation is
never realized practically due to the second term in
Eq.~(\ref{quant-rest-PSF}) which grows infinitely when $L\to\infty$. Thus,
Eq.~(\ref{quant-rest-PSF}) is a good illustration of the statement that
the ultimate limit of super-resolution in the reconstructed object is
given not by diffraction but by the quantum fluctuations of light
represented by the second term.

The number $L$ of terms in the sum (\ref{quant-PSF}) which determines the
width of the reconstruction PSF, depends on the signal-to-noise ratio in
the input object. To obtain the maximum $L$ we shall compare the
signal-to-noise ratio in the input object to that in the reconstructed
object. As follows from Eq.~(\ref{quant-rest-PSF}) with increasing $L$ the
signal-to-noise ratio in the reconstructed object deteriorates. We shall
assume that reconstruction of the object is possible until the limit when
the signal-to-noise ratio in the reconstructed object becomes unity.

Let us define the singal-to-noise ratio in the input object
as~\cite{Kolobov95}
\begin{equation}
    R = \frac{\langle {\hat N}\rangle^2}{\langle
    (\Delta {\hat N})^2\rangle},
         \label{SNR-IN}
 \end{equation}
where
 \begin{equation}
    \langle {\hat N}\rangle =  \int_{-1}^1
    \langle {\hat a}^{\dag}(s){\hat a}(s)\rangle ds,
 \end{equation}
is the total mean number of photons in the input object, and $\langle
(\Delta {\hat N})^2\rangle$ - its variance. Similarly we define the
signal-to-noise ratio $R^{(r)}$ in the reconstructed object as
\begin{equation}
    R^{(r)} = \frac{\langle {\hat N^{(r)}}\rangle^2}{\langle
    (\Delta {\hat N^{(r)}})^2\rangle},
         \label{SNR-OUT}
 \end{equation}
where the mean number of photons in the reconstructed object is given by
 \begin{equation}
    \langle {\hat N^{(r)}}\rangle =  \int_{-1}^1
    \langle {\hat a}^{(r) \dag}(s){\hat a^{(r)}}(s)\rangle ds.
 \end{equation}
The deterioration of the signal-to-noise ratio in the reconstructed object
can be described by the noise figure $F$,
 \begin{equation}
    F = \frac{R}{R^{(r)}},
 \end{equation}
that is commonly used in the literature about amplifiers. Because the
signal-to-noise ratio $R^{(r)}$ in the reconstructed object is always
smaller than that in the input object, the noise figure is always larger
than unity. If we assume that the minimum value of $R^{(r)}$ that allows
for reconstruction of the object is unity, this gives us the maximum noise
figure $F_{\rm max} = R$ corresponding to the maximum super-resolution.

Let us consider an input object in a coherent state, so that $\langle
{\hat a}(s)\rangle = a(s)$, $\langle {\hat a}^{\dag}(s){\hat a}(s)\rangle
= |a(s)|^2$, $\langle {\hat a}^{\dag}(s){\hat a}^{\dag}(s'){\hat
a}(s'){\hat a}(s)\rangle = |a(s)|^2|a(s')|^2$. It is easy to show that in
this case the input signal-to-noise ratio $R$ is equal to the mean total
photon number in the input object,
 \begin{equation}
    R = \langle {\hat N}\rangle =  \int_{-1}^1
    |a(s)|^2 ds.
 \end{equation}
On the other hand, for the signal-to-noise ratio $R^{(r)}$ in the
reconstructed object in this case we obtain the following result
 \begin{equation}
    R^{(r)} = \Bigl(\sum_{k=0}^{L-1}|a_k|^2\Bigr)^2/
    \Bigl(\sum_{k=0}^{L-1}\frac{|a_k|^2}{\lambda_k}\Bigr),
          \label{reconst_SNR}
 \end{equation}
where $a_k$ are the coefficients of decomposition of $a^{(r)}(s)$ over the
prolate functions $\varphi_k(s)$ in Eq.~(\ref{obj_reconst}).

As follows from Eq.~(\ref{reconst_SNR}), the signal-to-noise ratio
$R^{(r)}$ and, therefore, the noise figure $F$ depend on the shape of the
input object. For numerical evaluation of the super-resolution factor as a
function of the total mean number of photons in the input object we have
taken a narrow rectangular object placed at the origin $s=0$,
\BE
    a_{\epsilon}(s) = \left\{\begin{array}{cl}\displaystyle{
    \sqrt{\frac{\langle {\hat N}\rangle}{\epsilon}}} & \quad  |s|\leq \epsilon/2,\\
    0 & \quad  |s|>\epsilon/2,\\
    \end{array} \right.
 \EE
Taking the width $\epsilon$ of this object ever smaller we arrive at a
point-like source, while keeping the total number of photons constant and
equal to $\langle {\hat N}\rangle$. Such a point-like object gives us the
reconstruction PSF $h^{(r)}(0,s)$ at the output.

The degree of super-resolution in the reconstructed object can be
characterized by the ratio of the width of the diffraction-limited imaging
PSF to the width of the reconstruction PSF. In Fig.~6 we have shown the
imaging PSF $h(s)$ and the reconstruction PSF $h^{(r)}(0,s)$ for $L=7$
normalized to unity at their maxima. To define the super-resolution factor
we shall introduce the half-widths $W$ and $W_L$ of these two PSF measured
at their half-maxima. Then we define the super-resolution factor $S$ as
the ratio of $W$ to $W_L$,
\begin{equation}
    S = \frac{W}{W_L}.
 \end{equation}
For the example given in Fig.~6 these half-widths are equal to $W=1.895$,
$W_L=0.252$, and $S=7.5$.

In Fig.~7 we have plotted the super-resolution factor $S$ as a function of
the total mean number of photons $\langle {\hat N}\rangle$ in the input
object for the case of coherent light and multimode squeezed light. As
seen from this figure, for the same mean number of photons multimode
squeezed light provides higher super-resolution that the coherent light.

\section{Discussion and prospectives}
\label{Discussions}

In this paper we have investigated analytically and numerically the
quantum limits of super-resolution in reconstruction of optical objects
from the diffraction-limited images. We assume that such a reconstruction
is performed electronically from the data collected in a homodyne
detection of images by means of a CCD camera placed either in the image
plane or in the Fourier plane. We call the first situation {\it
super-resolving microscopy} and the second - {\it super-resolving
Fourier-microscopy}. In Section II we have presented the quantum theory of
the latter.

From our numerical simulations in Sections III and IV we conclude that
{\it a priori} information about the finite size of the input object
allows one to obtain significant super-resolution in the reconstructed
object over the classical Rayleigh limit. For classical noise-free objects
we have demonstrated a possibility to achieve super-resolution of factor 8
over the Rayleigh limit. When quantum fluctuations in the object are taken
into account, the degree of super-resolution depends on the
signal-to-noise ratio in the input object. When the object is illuminated
by the light in a coherent state, this signal-to-noise ratio is given by
the mean total number of photons during the observation time. We have
quantitatively defined the super-resolution factor as a ratio of widths of
the imaging and reconstruction point-spread functions. We have numerically
evaluated this super-resolution factor as a function of the mean total
number of photons for the coherent light and multimode squeezed light. For
the same mean number of photons multimode squeezed light allows to achieve
higher super-resolution than the coherent light.

As future prospectives for further development of the theory presented in
this paper we would like to mention the problem of read-out of binary
information from optical discs, like CD and DVD, or optical memory. The
storage density for the optical discs currently is limited by the spot
size of the diffraction-limited focussed light beam. One possibility of
increasing the storage density would be an attempt to put several bits of
information inside the diffraction-limited light spot with subsequent
super-resolution in the read-out of this information. In this case the
amount of {\it a priori} information is clearly superior to the case
considered in present article since for finite number of bits inside the
laser spot one has a finite possible combination of light patterns.
Therefore, one would expect higher potential of super-resolution in
read-out of binary information as compared to the situation that we have
investigated here. In the context of the problem of read-out of optical
discs one would have to generalize our theory to optical systems with high
numerical apertures which is usually the case for the systems of the
optical data storage.

This work was supported by the Project QUANTIM (IST-200-26019) of the
European Union.

\newpage

 \begin{figure}[t]
 \epsfxsize=15cm
 \centerline{\epsfbox{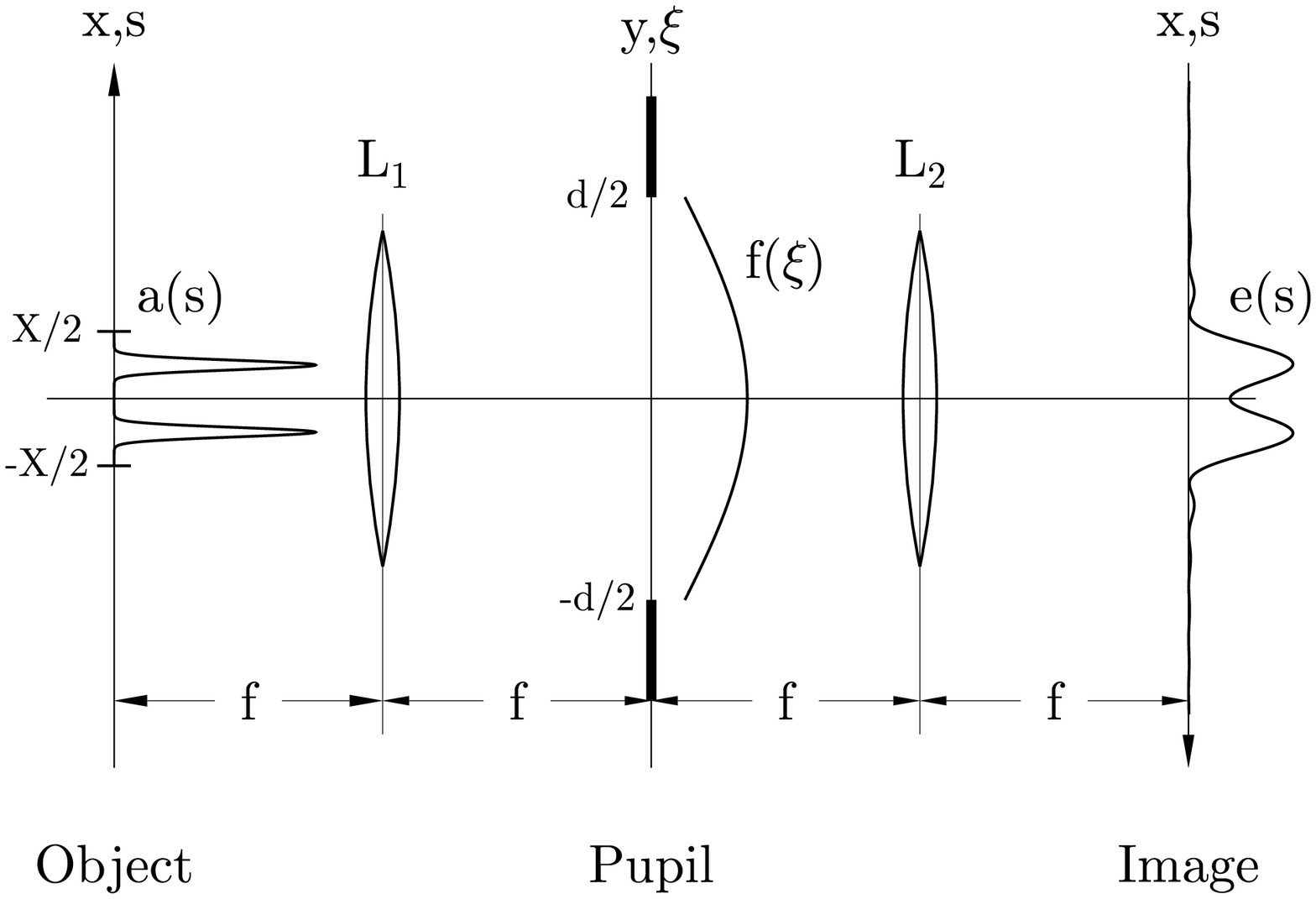}}
 \caption{Optical scheme of one-dimensional coherent diffraction-limited imaging.}
 \label{scheme}
 \end{figure}

 \begin{figure}[t]
 \epsfxsize15cm
 \epsfbox{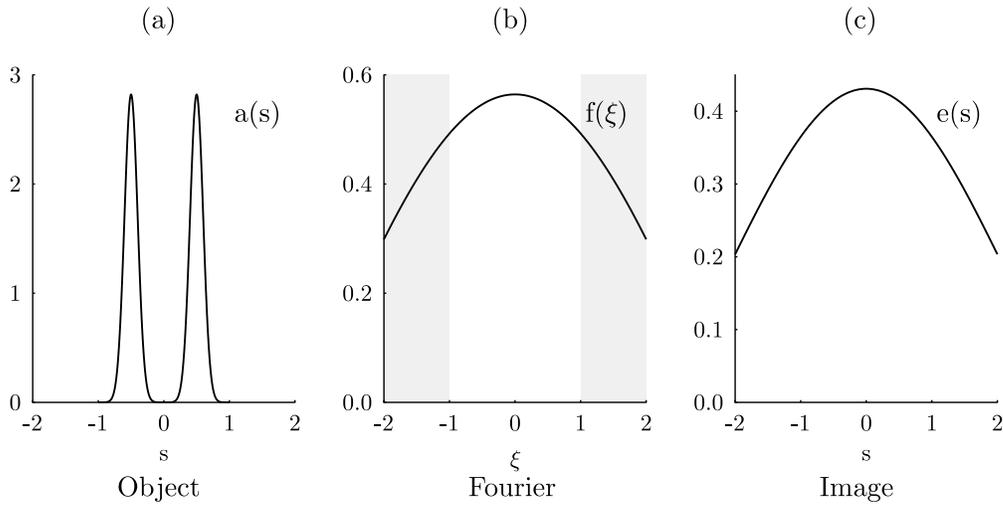}
 \caption{Double-peak object $a(s)$  used in numerical
simulations (a), its spatial Fourier spectrum $f(\xi)$ in the pupil plane
(b), and the image $e(s)$ created in the image plane (c). Grey area shows
the part of the spatial spectrum absorbed by the opaque area of the
pupil.}
 \label{fig:obj-img}
 \end{figure}

 \begin{figure}[t]
 \epsfxsize15cm
 \epsfbox{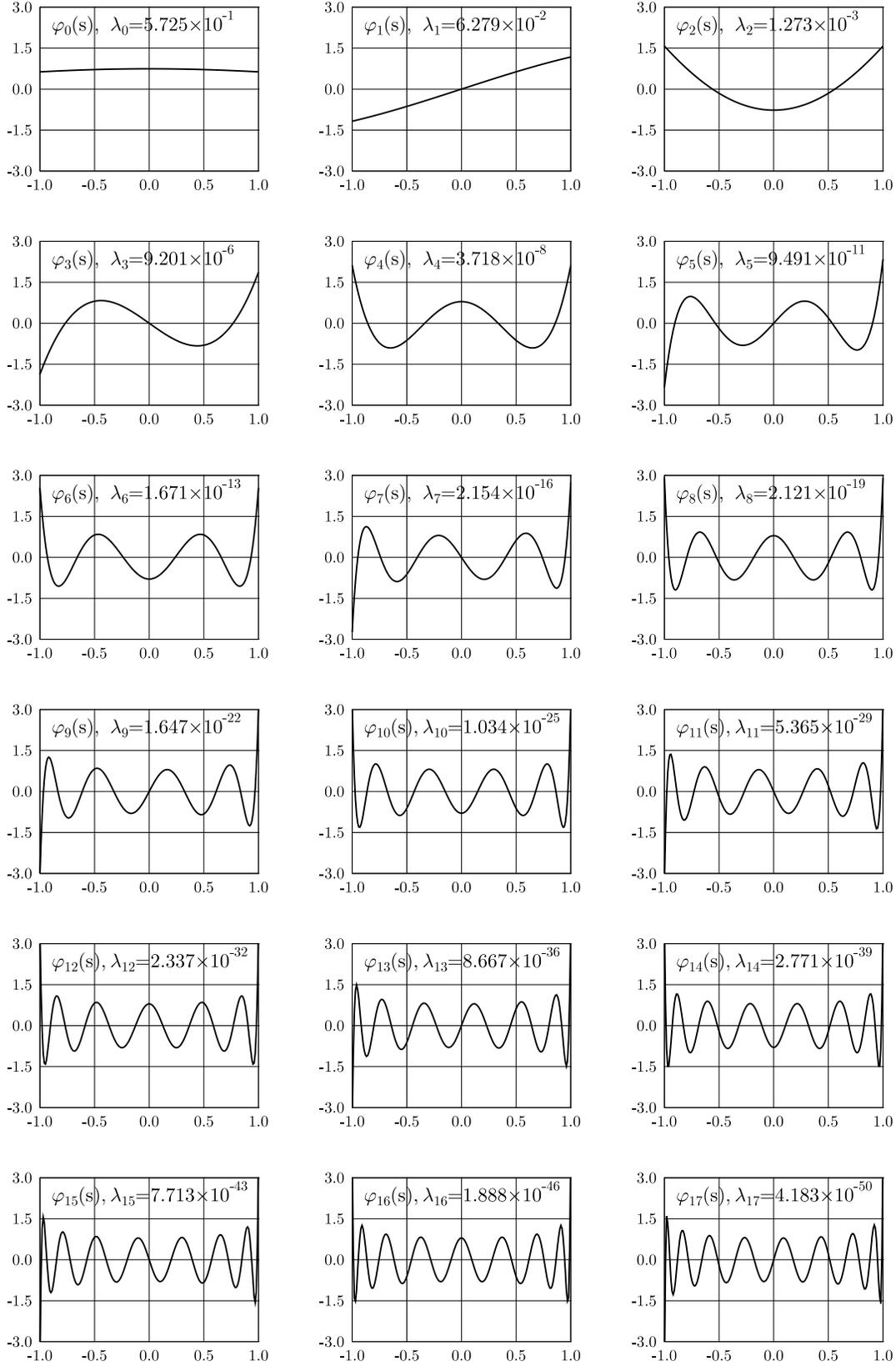}
 \caption{Examples of the prolate spheroidal functions $\varphi_k(s)$ and
 the corresponding eigenvalues $\lambda_k$ calculated using our numerical
 program.}
 \label{examples}
 \end{figure}

 \begin{figure}[t]
 \epsfxsize15cm
 \epsfbox{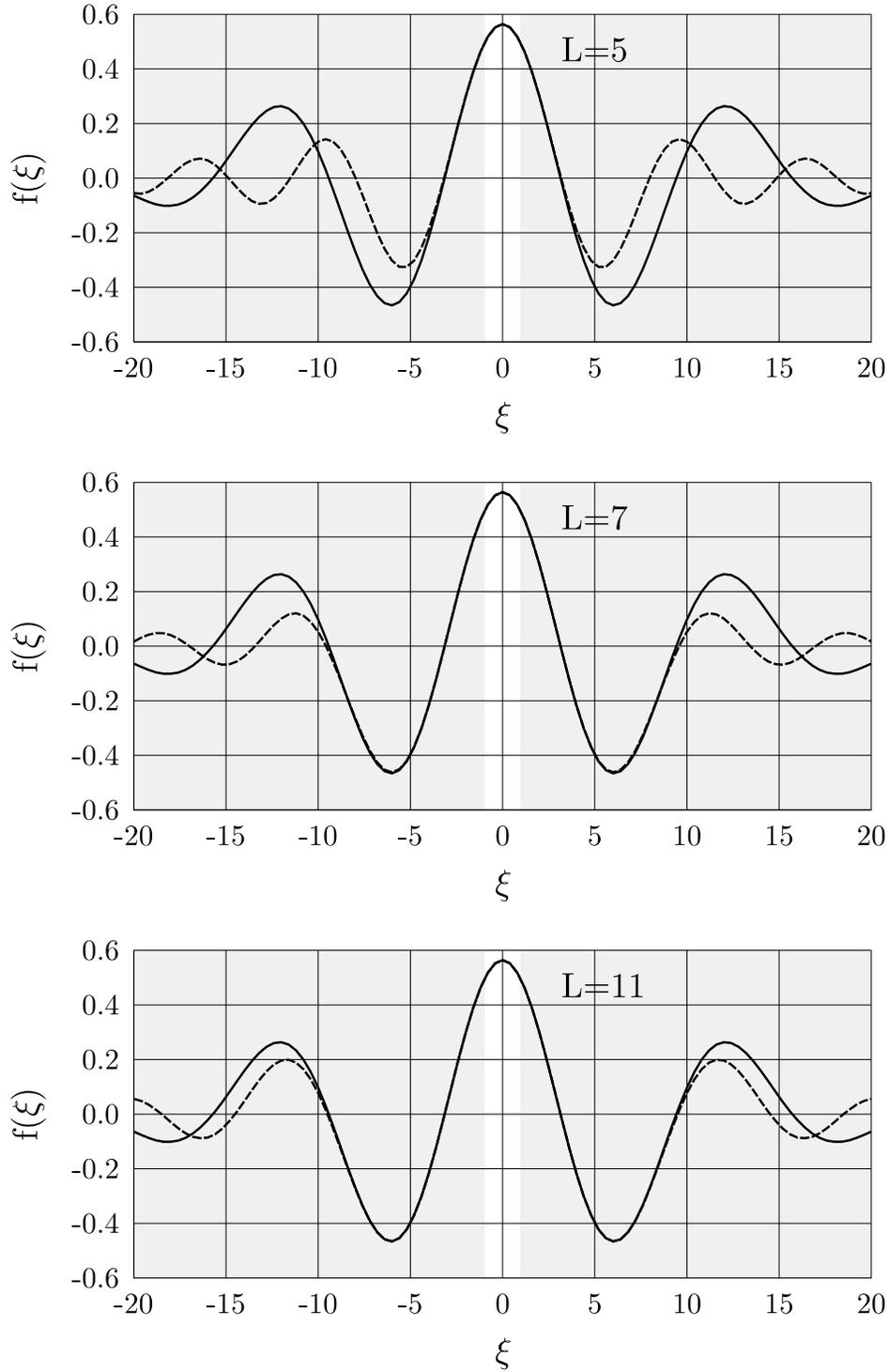}
 \caption{Exact spatial Fourier spectrum of the object from Fig.~2a (solid line),
 and the spectra reconstructed with $L=5, 7$ and 11 prolate functions
(dashed lines). Here and in Fig.~5 grey area indicates the absorbed part
of the spatial spectrum.}
 \label{fig:spectra}
 \end{figure}

 \begin{figure}[t]
 \epsfxsize15cm
 \epsfbox{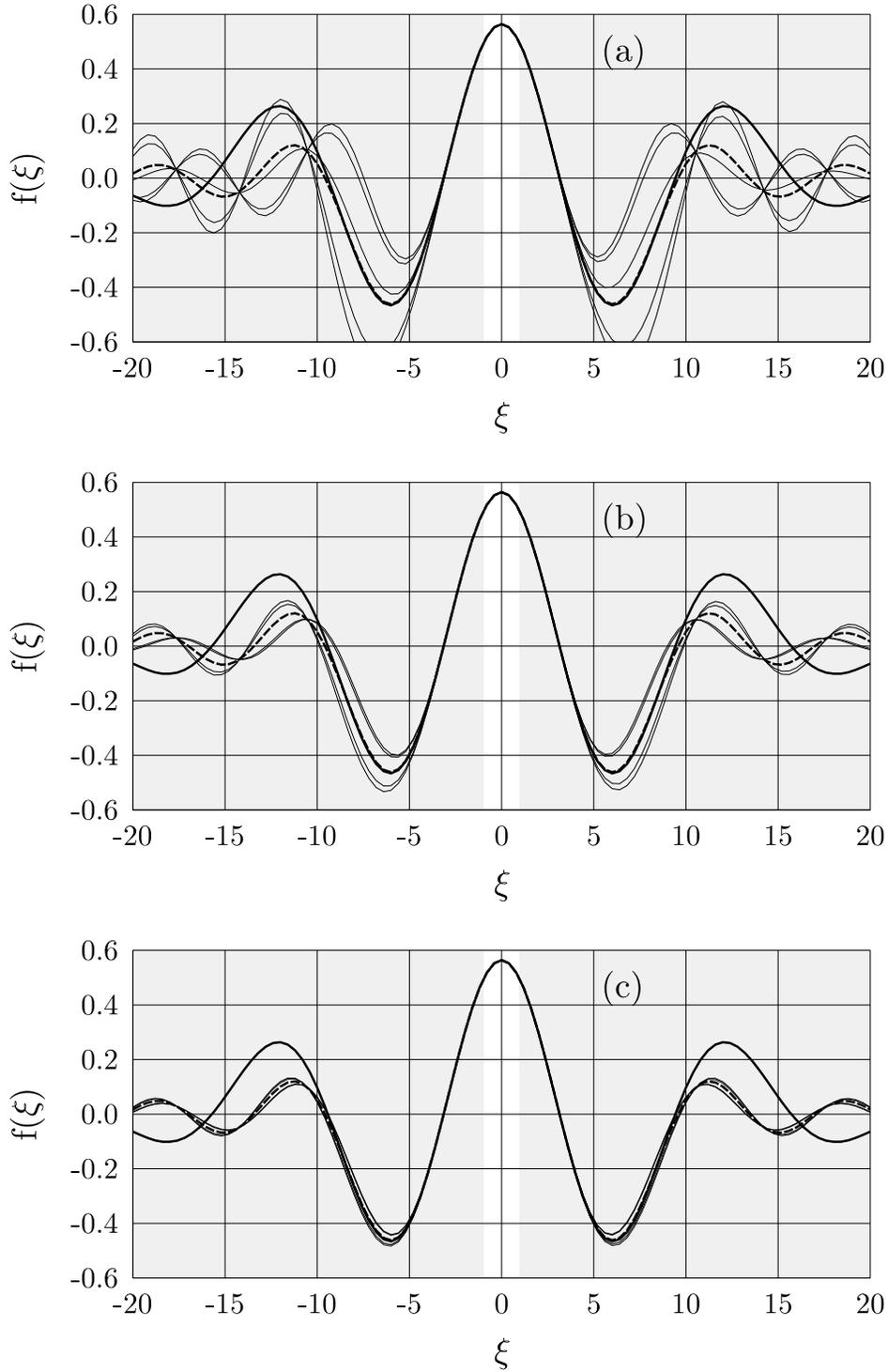}
 \caption{Exact spatial Fourier spectrum of the object from Fig.~2a (solid line),
 the reconstructed spectrum with $L=7$ prolate functions (dashed lines), and five
 random Gaussian realizations of the reconstructed spectrum with $L=7$ prolate functions
 (thin lines); (a) coherent light with mean total photon number
 $\langle {\hat N}\rangle = 10^{12}$, (b) coherent light with $\langle {\hat N}\rangle = 10^{13}$, and
 (c) squeezed light with $\langle {\hat N}\rangle = 10^{12}$ and $\exp(r)=10$.}
 \label{fig:spectra+fluctuations}
 \end{figure}

 \begin{figure}[t]
 \epsfxsize15cm
 \epsfbox{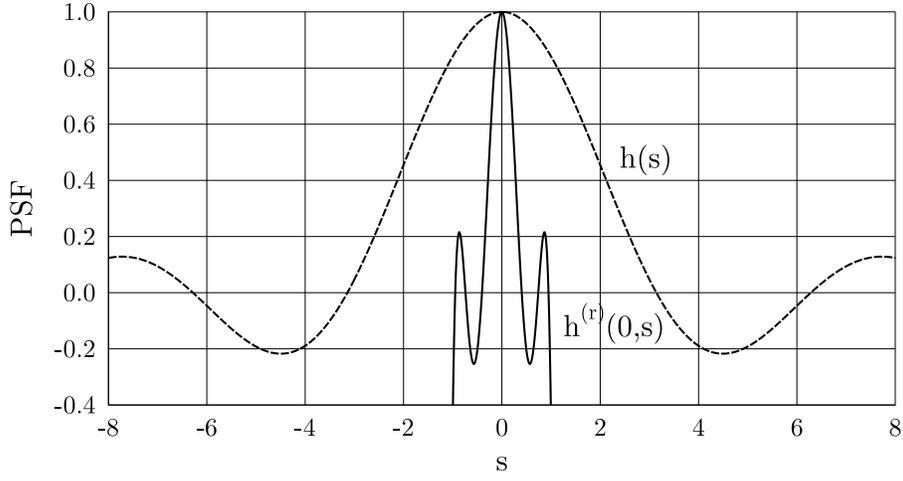}
 \caption{Diffraction-limited imaging point-spread function $h(s)$ and
 the reconstruction point-spread function $h^{(r)}(0,s)$ using 7 prolate
 functions.}
 \label{fig:point-spread_function}
 \end{figure}

 \begin{figure}[t]
 \epsfxsize15cm
 \epsfbox{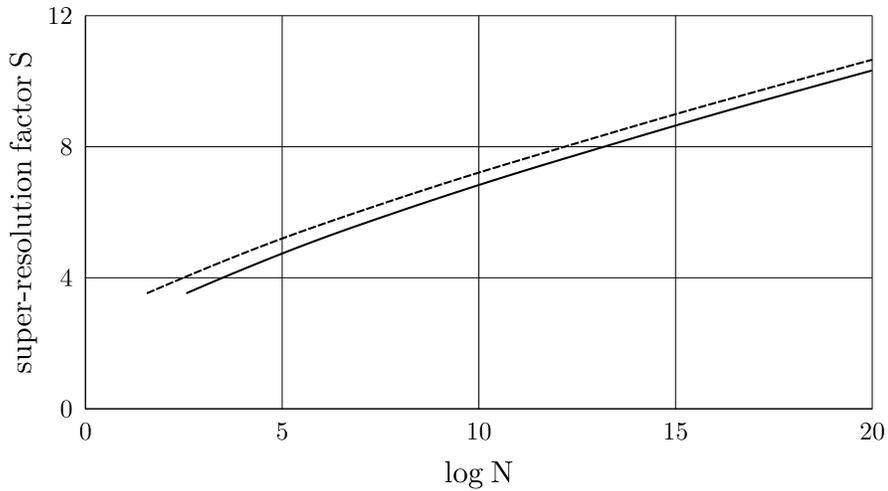}
 \caption{Super-resolution factor $S$ as a function of the total mean number of photons
 $\langle {\hat N}\rangle$ for coherent light (solid line, and multimode squeezed line
 with $\exp(r)=10$ (dashed line).}
 \label{fig:super-resolution}
 \end{figure}
\vfill

\end{document}